\documentclass[journal]{IEEEtran}

\ifCLASSINFOpdf
\else
   \usepackage[dvips]{graphicx}
\fi
\usepackage{url}
\usepackage{amsmath}
\usepackage{graphicx}
\usepackage{placeins}
\usepackage{float}
\usepackage{bbm}
\usepackage{subfigure}
\usepackage{verbatim}
\usepackage{amssymb}
\usepackage{amsmath}
\usepackage{amsthm}
\usepackage{mwe}

\usepackage{graphicx}

\begin{document}

\title{Multi-player Bandits for Distributed Cognitive Radar}

\author{William W. Howard, Charles E. Thornton, Anthony F. Martone, R. Michael Buehrer
\thanks{W.W. Howard, C.E. Thornton, and R.M. Buehrer are with Wireless@VT, Bradley Department of ECE, Virginia Tech, Blacksburg, VA, 24061. 
(e-mails:$\{$wwhoward, thorntonc, buehrer$\}$@vt.edu).  }
\thanks{A.F. Martone is with the U.S. Army Research Laboratory, Adelphi, MD 20783. (e-mail:anthony.f.martone.civ@mail.mil).}
\thanks{The support of the U.S. Army Research Office (ARO) is gratefully acknowledged. }\vspace{-.6cm}}

\maketitle
\pagenumbering{gobble}

\begin{abstract}
With new applications for radar networks such as automotive control or indoor localization, the need for spectrum sharing and general interoperability is expected to rise. 
This paper describes the application of multi-player bandit algorithms for waveform selection to a distributed cognitive radar network that must coexist with a communications system. 
Specifically, we make the assumption that radar nodes in the network have no dedicated communication channel. 
As we will discuss later, nodes can communicate \textit{indirectly} by taking actions which intentionally interfere with other nodes and observing the resulting collisions. 
The radar nodes attempt to optimize their own spectrum utilization while avoiding collisions, not only with each other, but with the communications system. 
The communications system is assumed to statically occupy some subset of the bands available to the radar network. 
First, we examine models that assume each node experiences equivalent channel conditions, and later examine a model that relaxes this assumption. 
\end{abstract}

\begin{IEEEkeywords}
radar networks, multi-arm-bandit, cognitive radar, reinforcement learning
\end{IEEEkeywords}
\IEEEpeerreviewmaketitle
\vspace{-.37cm}
\section{Introduction}
\vspace{-.1cm}

With the advent of fifth-generation (5G) cellular technologies, a large increase in the demand for spectrum access is expected. 
Since this will have an effect on all radio access devices, governing and standards agencies have a motivation to develop and implement access schemes that are not only robust to coexistence, but provide ever greater performance \cite{6967722}. 
With this drive for increased coexistence, cognitive and metacognitive strategies have been proposed as methods of enhancing the spectrum sharing capabilities of radar systems \cite{8000669}. 
This can be accomplished by adding spectrum sensing capabilities to radar systems, so that the system is able to build a characterization of its electromagnetic environment in real time \cite{9114775}. 

With the flexibility and agility offered by cognitive radar systems, reinforcement learning techniques have been implemented to address decision making in these systems \cite{9178313}. 
These approaches usually rely on  a history of spectrum activity and radar behavior, as well as an assumption that the environment obeys the Markov property. 
However, the Markov assumption is not always valid for realistic channels, such as when the interference has extended temporal correlations.

Because this assumption is not always justified, the probabilistic model of a \textit{multi-armed bandit} (MAB) is often considered for radar coexistence. 
The MAB model is a framework for sequential decision making problems, which involve a player seeking to choose the best action out of a set of actions over a series of time steps. 
The name originates from gambling on slot machines, which at one point were colloquially known as ``one-armed bandits.'' 
This framework has proved useful in numerous applications, including economic modeling \cite{10.5555/2832249.2832384}, healthcare \cite{pmlr-v85-durand18a}, and wireless communications \cite{7867071, 8190862}. 

MABs are useful due to their low complexity and robust theoretical guarantees \cite{MAL-068}. 
To address the problem of \textit{distributed} learning for radar waveform selection, when multiple agents seek to learn the best actions in a coordinated\footnote{Specifically by saying coordinated, we mean that each node knows the algorithm that each other node is using, but does not know the observations or decisions of other nodes. } manner \cite{NIPS2018_7952}, the MAB structure can be extended to a distributed learning scheme in the form of a multi-player MAB (or MMAB) \cite{6483307}. 

\textit{Contributions}. Cognitive strategies for radar networks have been studied before. 
The work of \cite{8048022} and \cite{7874202} study the power allocation problem and \cite{8464057} investigates the impact of node placement on interference. 
The performance of MAB algorithms in single-radar scenarios has also been investigated \cite{thornton2020efficient}. 
Our work applies several MMAB algorithms to the problem of coexistence between a cellular communication system and a cognitive radar network. 
We demonstrate that the proposed algorithm attains lower regret than comparable algorithms, where regret corresponds to cumulative instances of interference both with the communications system and between nodes. 
To the best of the authors' knowledge, this is the first work that applies the MMAB framework to the problem of distributed radar spectrum sharing. 


\section{Background} 

\subsection{Cognitive radar networks}
A \textit{cognitive radar} system is able to dynamically adapt parameters in response to feedback between the transmitted and received waveforms. 
This is typically accomplished via spectrum sensing, and has been proposed as a method for coexistence \cite{9114775, 8961364}. 
In contrast with work that seeks to adapt radar parameters to specifically improve target tracking performance, this work focuses on adapting parameters to accommodate for coexistence and mitigate interference.

A \textit{cognitive radar network}, then, is a collection of cognitive radars that collaborate to optimize any parameter of interest \cite{6586147, 1574168}. 
Thus, this work builds on research from two areas: (a) cognitive radar and (b) radar networks. 
The spectrum optimization problem for radar has been previously addressed in \cite{8360535}, from a radar node placement perspective. 
In this work we assume that the nodes are already placed and must learn a spectrum allocation. 
We're interested in scenarios which minimize mutual interference, so ideally each node utilizes a unique frequency band. 
Instead of relying on centralized coordination as is the case in some distributed MIMO radar systems, we assume that the cognitive strategies are implemented at each node in the network and that there is no dedicated communication channel between nodes. 
This allows for each node's choices to be made independently of every other node, and places no assumptions on the presence of a communications channel or even the number of nodes. 

It is assumed in this work that different spectrum bands are of different quality. 
In order to determine spectrum quality, we need to develop a metric. Then, radar nodes can evaluate the quality of a channel with given bandwidth and center frequency. 

We can represent the spectrum of any nonideal channel \cite{radar_channel_quality} with bandwidth $B_k$ and center frequency $f_k$ as
\begin{equation}
    \label{eq:nonideal}
    H_k(f) = A_k(f)e^{j\phi_k(f)}\text{rect}\left(\frac{f-f_k}{B_k}\right)
\end{equation}
where
\begin{align*}
    k &: \text{Channel index, some natural number}\\
    A_k(f) &: \text{A positive amplitude function in $\mathbb{R}$ for channel $k$}\\
    \phi_k(f) &: \text{A positive phase function in $\mathbb{R}$ for channel $k$}\\
    f_k &: \text{Center frequency for channel $k$}\\
    B_k &: \text{Bandwidth for channel $k$}\\
\end{align*}
Then we can represent an ideal channel also with bandwidth $B_k$ and center frequency $f_k$ as
\begin{equation}
    \label{eq:ideal}
    H_{\text{ideal}}(f) = A_{\text{ideal}}e^{j\phi_{\text{ideal}}(f)}\text{rect}\left(\frac{f-f_k}{B_k}\right)
\end{equation}
where $A_{\text{ideal}}$ is now a constant gain across the channel and $\phi_{\text{ideal}}(f) = -2\pi T_{\text{ideal}} f$ is the linear phase across the channel. 

We can measure the difference between any two channels $H_k, H_l$ as the Euclidean distance (or $L^2$ norm). 
Specifically we can compare any channel $H_k$ against an ideal channel (with unit gain) as 
\begin{equation}
    \label{eq:channel_comparison}
    Q_{k,l} =  \left[\frac{G_{k,\text{ideal}}}{M_{k,\text{ideal}}}\right]\text{Re}\{\rho_{k,\text{ideal}}\}
\end{equation}
where $G_{k,l}$ and $M_{k,l}$ are the geometric and arithmetic means, $\text{Re}\{\cdot\}$ is the real part of the argument, and $\rho$ is the correlation coefficient.  
\begin{equation}
    \label{eq:geo_mean}
    G_{k,l} = \sqrt{\langle|H_k(f)|^2\rangle\langle|H_l(f)|^2\rangle}
\end{equation}
\begin{equation}
    \label{eq:arith_mean}
    M_{k,l} = \frac{\langle|H_k(f)|^2\rangle + \langle|H_l(f)|^2\rangle}{2}
\end{equation}
\begin{equation}
    \label{eq:coherence}
    \rho_{k,l} = \frac{\langle H_k(f)\overline{H}_l(f)\rangle}{\sqrt{\langle|H_k(f)|^2\rangle\langle|H_l(f)|^2\rangle}}
\end{equation}
Here, $\langle \rangle$ denotes the mean value and $Q$ falls in $[0,1]$.

\subsection{Multi-arm bandits}
The multi-arm bandit literature focuses on repeated interactions between one or more \textit{players} and an \textit{environment}. 
Generally, time is divided into intervals with some finite horizon during which each player chooses to play one of several available arms (i.e., actions), observes a reward, and is informed whether any other players selected the same arm. 
In wireless communications applications, the actions are usually represented by a vector of possible transmission parameters. 
Each arm has an associated reward drawn i.i.d. from an unknown distribution. 
At the start of the game, the only information available to the players is the number of available arms - they do not know how many other players there are, or the mean rewards for each arm. 

Considering the distributed radar problem, the set of nodes are players, the options for channels become arms, and the shared spectrum is the environment. 

\textit{Notation}. 
Script capitol letters $\mathcal{D}$ represent sets, and the same letter in lower case $d$ represents an element. 

Let $\mathcal{P}$ be the set of players and $\mathcal{A}$ be the set of actions. 
With finite horizon $T$, in time step $t \in \mathbb{N}, t \leq T$, player $n\in\mathcal{P}$ selects the $i^{th}$ action denoted as $A_{i,n}(t)$. 
The player then observes the corresponding reward $R_{i,n}(t)$ which is drawn i.i.d. from an unknown distribution\footnote{In the literature, this is typically Gaussian or Bernoulli, but no assumption on the reward distribution is needed. } with mean $\mu_{i,n}$ and falls in the closed interval $[0,1]$. 

The sets $\mathcal{P}$ and $\mathcal{A}$ form the two parts of a bipartite graph. 
An \textit{edge} is a connection between node $p_i$ and action $a_i$ denoted simply as $p_i a_i$. 

We can define a \textit{mapping} $\mathcal{E}$ between $\mathcal{P}$ and $\mathcal{A}$ as any set of edges $p_i a_i$ which each connect one node to one sub-band.  Possibly, more than one player can be mapped to the same sub-band. If this is the case, we say a \textit{collision} has occurred and the affected players observe a reward of 0. 

Now, a \textit{matching} between the set of players $\mathcal{P}$ and the set of arms $\mathcal{A}$ is any one-to-one mapping, denoted as $\pi: \mathcal{P} \to \mathcal{A}$. 
Matchings are characterized as not containing any collisions. 
The $\textit{utility}$ of a matching $\pi$ is $U(\pi)= \sum_{n=1}^P \mu_{n,\pi(n)}$. 
The set $\mathcal{M}$ contains all possible matchings, and $U^* = \max_{\pi \in \mathcal(M)}U(\pi)$ is the maximum utility over all possible matchings. 

Each player does not observe any other player's selections, other than observing whether a collision has occurred. Players observe the output of an indicator function of whether their edge shares an endpoint with any other player (i.e., of whether a collision occurs). If we call $\mathcal{E}'$ the set colliding edges in $\mathcal{E}$, then players observe an indicator of whether their edge $p_i a_i$ is in $\mathcal{E}'$. 
\begin{equation}
\label{eq:collision_indicator}
    \mathbbm{1}_{\mathcal{E}'}(i) = \begin{cases}
    1, & \text{if\;\;} p_i a_i \in \mathcal{E}' \\
    0, & \text{else} \\
    \end{cases}
\end{equation}
Practically, this can be implemented as an SINR threshold $\eta$. If the SINR drops below this threshold then it can be assumed that more than one radar node collided, or that the radar collided with a communications system. 

Each radar node only observes indicators and rewards for the actions they choose at each step. 
Over time, the player is able to estimate the mean rewards for each action. 

After each observation, each player updates the average observed reward for the arm they selected. 
A simple strategy might have each player try each arm once, check which one has the highest observed reward, and play that arm for the rest of the game. 
However, this strategy is naive, since the player will not have a very good characterization of the mean arm rewards from a single experience alone. 
So, a better strategy would have each player balance actions that \textit{explore} (better characterize the environment) and actions that \textit{exploit} (favor high expected reward, where the expectation is over past observations). 

When we consider multiple players, it is assumed that they are playing on the same bandit arms (in other words, they must share the same set of sub-bands). 
Since each player must choose an arm in each time step, it is possible that more than one will select the same arm and therefore interfere with each other, which provides a reward of 0. 
Since each player's goal is to maximize its expected reward, colliding is not favorable. 

Each node's different physical location and possibly different array characteristics might lead to a different observed sub-band quality. 
Therefore, we assume that the mean reward for each action and each player can be different, or in other words that it has a different channel quality as measured in (\ref{eq:channel_comparison}). 

We consider a fully distributed learning scheme where there is no fusion center and no dedicated communication channel between nodes. 
Therefore, there can be no centralized scheduling or pre-allocated frequencies for each node; therefore, they each make their own decisions. 
This contrasts with the scenario where a central decision maker coordinates the actions that each node selects through some algorithm.  

\section{Modeling}
Let the cognitive radar network contain $P$ nodes which share $S > P$ equally sized sub-bands, one or more of which is statically occupied by a communications system. 
Time is divided into slots, starting at $t=0$ for all players and ending at finite horizon $T$. 
In each slot (or pulse repetition interval (PRI)), each node learns from it's previous actions, selects and transmits a Linear Frequency-Modulated (LFM) chirp waveform in one of the sub-bands, and senses the spectrum for collisions. 

Each waveform is selected from a finite collection $\mathcal{W} = \{w_i\}_{i=1}^W$, and is given by \cite{handbook}
\begin{equation}
    w_i(t) = A \; \text{rect}\left(\frac{t}{T}\right) \cos{(2\pi f_{c,i}t + \pi \alpha t^2)}
\end{equation}
where $t$ is continuous time, $A$ is a constant amplitude, $T$ is the pulse duration, $f_{c,i}$ is the center frequency, and $\alpha_i$ is the up-chirp rate which corresponds to the bandwidth as $Bw_i = T\alpha_i$. 
The network then seeks to learn a sub-band allocation for each node that respects the presence of the communications system, and allows for minimal collisions. 

Each reward $R_{i,n}(t)$ is drawn i.i.d. from some distribution $\mathnormal{f}$ with mean $\mu_{i,n}$. 
For our application, this random draw makes sense since the channel quality will vary with time due to the random electromagnetic environment. 
In the \textit{homogeneous} setting, the vector of mean rewards for each action and each player can be represented as $\overline{\mu}_{n,k} = \overline{\mu}_{k}$. 
In other words, each player's rewards are drawn with the same mean. 
Then, in the \textit{heterogeneous} setting, the vector of mean rewards for each player can differ: 
\begin{equation}
    \label{eq:mean_matrix}
    \overline{\mu} = 
    \begin{bmatrix}
    \mu_{1,1}  & \mu_{1,2}  & \cdots & \cdots & \mu_{1,S} \\
    \mu_{2,1}  & \ddots  & \ddots && \vdots \\
    \vdots & \ddots & &  & \vdots \\
    \vdots &  &  & & \vdots \\
    \mu_{P,1} & \ddots & \ddots & \ddots & \mu_{P,S}\\
    \end{bmatrix}
\end{equation}
In our context, the mean reward is based on the channel quality as observed by each node. 
Specifically, with sub-band $i$ as sensed by player $n$, using (\ref{eq:channel_comparison}), the mean reward is calculated as
\begin{equation}
    \label{eq:pre-reward}
    \hat{\mu}_{i,n} = Q_{(i,n), ideal}
\end{equation}
In order to use the entire interval $[0,1]$ we can normalize the mean reward vector by dividing by the quality of the best sub-band: 
\begin{equation}
    \label{eq:reward_vec}
    \overline{\mu}_n = \frac{[\hat{\mu}_{n, 1}, \hat{\mu}_{n,2}, ..., \hat{\mu}_{n,S}]}{\max_k{(\hat{\mu}_{n,i}})}
\end{equation}
This is valid since the radar does not directly calculate the reward, it is generated by the environment. 

The presence of the communication system is represented as another indicator function, which is not observed by the radar nodes. Let $\mathcal{A}'$ be the set of sub-bands occupied by the communications system. We can summarize the reward function as
\begin{align}
    \label{eq:reward_fn}
    R^{i,n}(t) &= x_{i,n}(t)\mathbbm{1}_{\mathcal{E}'}^{(t)}(i)\mathbbm{1}_{\mathcal{A}'}^{(t)}(i)\\
    x_{i,n} &\sim \mathnormal{f}(\mu_{i,n})
\end{align}
where $\mathbbm{1}_{\mathcal{E}'}^{(t)}(i)$ is the radar collision indicator, and $\mathbbm{1}_{\mathcal{A}'}^{(t)}(i)$ indicates a collision with the communications system. 
In our simulations, $\mathnormal{f} \sim \mathcal{N}(\mu_{i,n}, \sigma)$ is normally distributed with mean depending on the node and sub-band and specified variance. 

We can then define the \textit{regret}\footnote{Regret, which comes in several varieties, is a measure of the quality of the players' actions, but is not observed or calculated by the players. It is used to analyze the overall performance.} more formally below. 
This is the ``weak regret", which compares the selected actions to the actions that are best on average. 
This metric corresponds to how well the spectrum is being utilized and is defined as:  
\begin{equation}
    \label{eq: regret}
    R_t = tU^* - \mathbb{E}\left[\sum_{t_0=1}^t\sum_{n=1}^P R^{i,n}(t_0)\right]
\end{equation}
at any time step $t$. 
We will investigate the cases when a unique optimal matching exists and when there may be more than one optimal matching. 
Note however that since the rewards reflect the channel quality in the environment as a real number in $[0,1]$ that there will very likely exist a unique optimal matching. 

Several algorithms have been presented in the literature to solve both the homogenous and heterogenous scenarios. 
Specifically we focus on four algorithms that have been developed for the multiplayer bandit problem: \textbf{Multiple UCB1} \cite{bandits}, \textbf{Musical Chairs} \cite{pmlr-v48-rosenski16}, \textbf{SIC} \cite{NIPS2019_9375}, and \textbf{M-Etc-Elim} \cite {mehrabian20a}. We detail these below. 

\textit{Multiple UCB1}. 
A simple implementation is to have each node implement a \textit{single}-player bandit model. 
Here we pick UCB1, and measure cumulative regret as the sum of each node's regret. 
Since this algorithm simply selects the best sub-band over time and implements no multiplayer strategies, there will be no intentional collisions. 
\begin{figure*}
    \centering
    \includegraphics[scale=0.55]{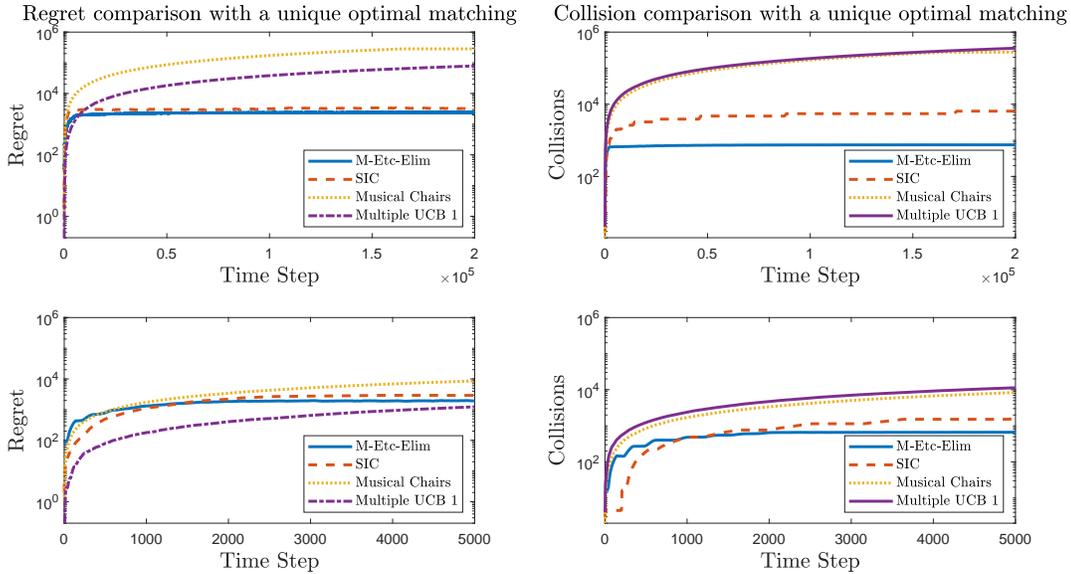}
    \caption{Comparison of four algorithms: Multiple UCB1, Musical Chairs, SIC, and M-Etc-Elim. On the right we see regret, and on the left are collisions. The top two plots show a full range of time steps, while the bottom two plots zoom in to view the early behavior of the algorithms. In each run, there is a single subset of the available bands that are optimal. }
    \label{fig:wombo}    
\end{figure*}

\textit{Musical Chairs}. 
This preliminary algorithm is the first one we present that is structured for multiple players. 
This means that it implements some structure that each player can implement to provide some coordination, even while they don't exchange information. 
In contrast with a later algorithm, Musical Chairs assumes that rewards are \textit{homogeneous}, i.e. $\overline{\mu_{n,k}} = \overline{\mu_k}$. 
To begin this algorithm, nodes explore randomly until with high probability, they learn a correct ranking of the estimated mean arm rewards as well as $P$, the number of nodes. 
Since the players observe the same average rewards, it is clear that they will agree on the subset of the best $P$ arms. 
The name "musical chairs" comes from the next part of this algorithm, wherein each player attempts to use a random arm in the set of best arms.
If they collide they try a new arm, but if they do not collide then they stay on that arm. 
This continues until all players are matched to a single arm in the set of best arms. 
At this point, the player will exploit this arm until the end of the game. 
The arm each player selects is referred to as that player's \textit{external rank}. 

\textit{SIC}. 
This algorithm was introduced to address the idea that any sort of synchronization between nodes will require communication, hence the name Synchronization Involves Communication. 
Rather than rely on an explicit, separate, communications channel or system, this algorithm introduces the notion of \textit{implicit communication} where players are able to exchange information through intentional collisions. 
Players use the above Musical Chairs algorithm to assign external ranks, then use another procedure to assign internal ranks. 
This is necessary since the players need to develop an agreed-upon ordering, and the external ranking might not be unique. 
While the external ranking corresponds to a single action (i.e., external ranks can be any value in $[1,2,\dots, A]$), the internal rankings take consecutive integers between $1$ and $P$. 

Following the initialization, players alternate between an \textit{exploration} phase and a \textit{communication} phase $m$ times. 
In the $m^{th}$ exploration phase, players follow the internal ranking to visit each arm $2^m$ times. 
Since the players are going in the order of their internal rankings, this phase is collision-free and lasts $P2^m$ time steps. 
In the $m^{th}$ communication phase, each player communicates its observed mean arm reward (truncated to $m+1$ bits) to all the other players. 
We will not go into detail regarding the communication algorithm other than to say that when player $n$ is communicating to player $l$, it pulls player $l$'s communicating arm to transmit a $1$ and it pulls its own communicating arm to transmit a $0$. 
When player $n$ is receiving information, it selects only its communication arm to allow collisions with transmitting players. 
This protocol incurs a lot of collisions, since each player communicates to each other player, which is a maximum of $P*(P-1)^{S*(m+1)}$ collisions in the $m^{th}$ communication phase.  


\textit{M-Etc-Elim}. 
This algorithm uses an assumption of \textit{heterogeneity} - i.e., arm reward means between players might vary. 
In a radar implementation, this assumption relates to the possibility of different channel conditions among the physical locations of the nodes. 
This scenario creates a much more challenging problem, especially since there are no dedicated communications between the radar nodes. 
The M-Etc-Elim algorithm builds on the idea of implicit communication introduced by SIC. 
In the initiation phase of this algorithm, nodes estimate their own arm means and $P$, then collaboratively assign external and internal ranks. 
Whichever node is ranked `1' becomes the \textit{leader}, with the rest becoming \textit{followers}. 
Through implicit communications, the follower nodes can transfer their observed mean sub-band rewards to the leader, who creates a graph of the possible matchings. Initially, all of the possible matchings are considered and as the game goes on this reduces to the set of optimal matchings. 
This method incurs less regret in each communication phase than in SIC, since rather than each player communication to each other player, the only communication is between the leader and the followers. 

If a unique optimal matching exists, then once the algorithm identifies it, each node exploits their action for the rest of the game. 
Otherwise, the nodes will explore between the multiple optimal matchings. 
Due to this, we see that the regret for this algorithm reaches a plateau then incurs zero regret for the remainder of the game. 

After the followers communicate their estimated mean action rewards, the leader compiles them into a matrix $E$ of the edge weights of the bipartite graph formed by the set of players and arms. 
Using some assignment algorithm (e.g. the Hungarian algorithm), the leader solves for the matching with the optimal utility $U^*$. 
In a general case, there may be more than one matching with utility $U^*$, but since the rewards are defined in $[0,1]$ it is unlikely that this will happen. 

\section{Simulations}

\subsection{Setting}
Let $P=4$ and $A=6$. The first sub-band is constantly occupied by a communications system, so we set the rewards for that sub-band to 0. Note that this is not the same as a five sub-band setting, since players can still use the first sub-band; they will just incur regret and collisions in doing so. 

Comparing all four of these algorithms directly is difficult, as some are meant for homogeneous scenarios and some for heterogeneous. 
Because of this, the comparisons we show are between scenarios appropriate for each algorithm. We show the specific values used in Eq. (\ref{eq:heterogeneous}) and Eq. (\ref{eq:homogeneous})
%
\begin{equation}
    \label{eq:heterogeneous}
    \overline{\mu} = 
    \begin{bmatrix}
    0  & 0.9  & 0.3 & 0.3 & 0.3 & 0.3 \\
    0  & 0.3  & 0.8 & 0.3 & 0.3 & 0.3 \\
    0  & 0.3  & 0.3 & 0.7 & 0.3 & 0.3 \\
    0  & 0.3  & 0.3 & 0.3 & 0.6 & 0.3 \\
    \end{bmatrix}
\end{equation}
Then, for the corresponding simulations with homogeneous rewards, we use 
\begin{equation}
    \label{eq:homogeneous}
    \overline{\mu} = 
    \begin{bmatrix}
    0  & 0.9  & 0.8 & 0.7 & 0.6 & 0.3 \\
    \end{bmatrix}
\end{equation}

In each case the variance is set as 0.01. Each time step, or radar PRI, is $0.1024ms$, and we simulate 200,000 PRI's. 
%

\subsection{Analysis} 
\begin{figure}[H]
    \centering
    \includegraphics[scale=0.55]{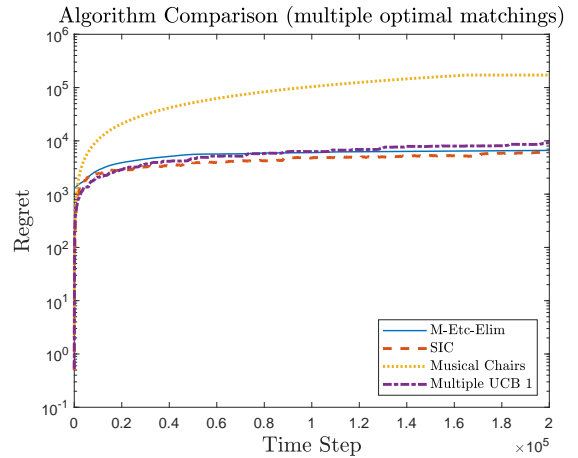}
    \caption{M-Etc-Elim vs Musical Chairs vs SIC under Gaussian rewards with more than one optimal matching}
    \label{fig:more_than_one}
\end{figure}

Since the M-Etc-Elim regret has a near-logarithmic upper bound, we anticipate that the regret converges rapidly and flattens in the exploitation phase. 
Fig. \ref{fig:wombo} demonstrates this nicely. We can see that the M-Etc-Elim algorithm experiences a lot of interference early in the game as the nodes exchange information through collisions. 
This is a good trade-off, since we can see that this algorithm achieves a lower regret bound than the other two multi-player algorithms. 
In Fig. \ref{fig:wombo}, we can also see the collision behavior of each algorithm. 
We would expect the Multiple UCB1 performance to be similar to that of Musical Chairs, since they neither of them implement much coordination. 
In time, Multiple UCB1 will learn to avoid the communication sub-band, since the reward there will always be less than the other sub-bands. 
We can see the effect of the longer exploration periods during the initiation of the game for all of the algorithms except for Multiple UCB1, causing the steeper regret for these. 
This quick convergence behavior is due to the fact that an \textit{optimal matching} exists, so the algorithm is able to identify that matching and settle on it for the remainder of the game. 
Since each time step, or PRI, lasts 0.1024ms and we can see that the M-Etc-Elim algorithm converges in around 1000 time steps, the convergence happens in about 0.1s. 
This allows the cumulative regret to become constant, since the radar nodes are not colliding with each other or with the communications system. 
To contrast this behavior, we can look at Fig. \ref{fig:more_than_one} where the optimal matching is not unique. 
While the regret over time still converges to a constant, we can see that the algorithm takes longer to converge since it has to identify and explore both optimal matchings.
Note that only the M-Etc-Elim performance is affected by the presence of more than one optimal matching. 
This is because this is the only algorithm that is able to recognize whether the optimal utility is unique or not. 

\section{Conclusion}
In this work we have shown that for a cognitive radar network with no explicit communications channel or system, using the M-Etc-Elim algorithm can guarantee a near-logarithmic regret bound in a relatively short amount of time. 
This is useful in scenarios where the network must share spectrum not only between its own nodes, but with communications systems that occupy some subset of the available spectrum. 

Given that implemented distributed radar systems often have some degree of communication, future work could extend this model to allow for limited exchange of information on a dedicated communication channel. In addition, since cognitive radar systems are able to sense the spectrum, this information can be used as context for decision making. Finally, different interference schemes could be time-varying and frequency agile. 
\bibliographystyle{IEEEtran}
\bibliography{bibli}

\end{document}